**Title:**
Natural selection reduced diversity on human Y chromosomes.

**Authors and affiliations:**
Melissa A. Wilson Sayres[1,2], Kirk E. Lohmueller[2,3,4], and Rasmus Nielsen[1,2]

1. Statistics Department, University of California-Berkeley, Berkeley, California 94720
2. Integrative Biology Department, University of California-Berkeley, Berkeley, California 94720
3. Department of Ecology and Evolutionary Biology, University of California, Los Angeles, CA, 90095.
4. Interdepartmental Program in Bioinformatics, University of California, Los Angeles, CA, 90095.

**Corresponding author:**
Please address correspondences to M.A. Wilson Sayres, mwilsonsayres@berkeley.edu

**Abstract**
The human Y chromosome exhibits surprisingly low levels of genetic diversity. This could result from neutral processes if the effective population size of males is reduced relative to females due to a higher variance in the number of offspring from males than from females. Alternatively, selection acting on new mutations, and affecting linked neutral sites, could reduce variability on the Y chromosome. Here, using genome-wide analyses of X, Y, autosomal and mitochondrial DNA, in combination with extensive population genetic simulations, we show that low observed Y chromosome variability is not consistent with a purely neutral model. Instead, we show that models of purifying selection are consistent with observed Y diversity. Further, the number of sites estimated to be under purifying selection greatly exceeds the number of Y-linked coding sites, suggesting the importance of the highly repetitive ampliconic regions. While we show that purifying selection removing deleterious mutations can explain the low diversity on the Y chromosome, we cannot exclude the possibility that positive selection acting on beneficial mutations could have also reduced diversity in linked neutral regions, and may have contributed to lowering human Y chromosome diversity. Because the functional significance of the ampliconic regions is poorly understood, our findings should motivate future research in this area.

**Author Summary**

The human Y chromosome is found only in males, and exhibits surprisingly low levels of genetic diversity. This low diversity could result from neutral processes, for example, if there are fewer males successfully mating (and thus fewer Y chromosomes being inherited) relative to the number of females who successfully mate. Alternatively, natural selection may act on mutations on the Y chromosome to reduce genetic diversity. Because there is no recombination across most of the Y chromosome all sites on the Y are effectively linked together. Thus, selection acting on any one site will affect all sites on the Y indirectly. Here, studying the X, Y, autosomal and mitochondrial DNA, in combination with population genetic simulations, we show that low observed Y chromosome variability is consistent with models of purifying selection removing deleterious mutations and linked variation, although positive selection may also be acting. We further infer that the number of sites affected by selection likely includes some proportion of the highly repetitive ampliconic regions on the Y. Because the functional significance of the ampliconic regions is poorly understood, our findings should motivate future research in this area.

**Introduction**

The Y chromosome has often been used as a marker for studying human demographic history [1], but one implicit assumption in these analyses is that the Y chromosome is not affected by natural selection at linked sites [2]. However, formal tests of models of selection have been lacking. In part, this has been due to a paucity of resequencing data for many male human genomes, where autosomal, X, Y and mtDNA for the same individuals could be compared. Such data eliminate one source of sampling variance that could influence comparisons between genomic regions, and also allow for chromosome-wide estimates of genetic diversity on the Y, which is often ignored in whole-genome analyses [3-5]. Under simple neutral models with constant and equal male and female population sizes, diversity is expected to be proportional to the relative number of each chromosome in the population: X diversity is expected to be three-quarters autosomal diversity (because there are three X chromosomes for every four autosomes) and both the Y and mtDNA diversity are expected to be one-quarter autosomal diversity [6].

The Y chromosome does not undergo homologous recombination, except in the small pseudoautosomal regions [7]. In general, diversity is reduced in genomic regions or genomes with little or no recombination [8-11]. Similarly, previous studies of small segments of the human Y chromosome have found low levels of genetic diversity, but multiple theories exist to explain this reduction [12-16].

Because the Y chromosome is found only in males, low diversity on the Y could result from neutral processes if, for example, the effective population size of males is reduced relative to that of females. One factor that can reduce the male population size is high variance in the number of offspring. Differences in the variance in reproductive success between the sexes, will cause differences in effective population sizes, even when the actual number of males and females is approximately the same [4,13]. Based on comparing patterns of genetic variation on the X chromosome and the autosomes, several recent studies have found evidence of sex-biased demographic processes during human history [3-5,17-20], often suggesting that the effective population size of females was higher than that of males throughout recent human history ($N_f > N_m$, if $N_f$ represents the effective number of breeding females and $N_m$ represents the effective number of breeding males).

Alternatively, purifying selection acting to remove new deleterious mutations on the Y chromosome, will affect diversity at linked neutral sites through a process called background selection. Background selection refers to the reduction in genetic diversity at sites that are themselves neutrally evolving, but are linked to other sites where deleterious mutations occur [21-24]. Background selection may be particularly potent on the Y chromosome, because there is no recombination on the Y chromosome. As such, deleterious mutations in one area of the chromosome could reduce levels of genetic diversity across the entire chromosome [12,14-16]. However, the strength of selection is also important. Several weakly deleterious mutations may interact resulting in a Hill-Robertson interference [25], whereby interference among linked sites weakens their effects on linked neutral sites [26]. Similarly, positive selection, acting on beneficial mutations is expected to decrease diversity at linked neutral sites. Given the unique gene content and lack of recombination on the Y chromosome, it is likely to have experienced a complex evolutionary history.

Here, using genome-wide analyses of X, Y, autosomal and mitochondrial DNA, in combination with extensive population genetic simulations, we show that low observed Y chromosome variability is not consistent with a purely neutral model. Instead, we show that models of purifying selection and background selection affecting linked neutral sites are consistent with observed Y diversity. Further, the number of sites estimated to be directly under purifying selection greatly exceeds the number of Y-linked coding sites, suggesting the importance of the highly repetitive ampliconic regions [27-29]. Because the functional significance of the ampliconic regions is poorly understood, our findings should motivate future research in this area.

**Results and Discussion**

**Diversity across the entire human Y is extremely low**

Analyzing complete genomic sequence data from 16 unrelated males (Table S1), we observe that normalized diversity on the human Y is extremely low compared to expectations from other genomic regions (Figure 1; Table 1). By analyzing resequencing data for the autosomes, X chromosome, Y chromosome, and mitochondria from the same individuals, we reduce sampling variance that might otherwise confound comparisons between regions of the genome. Here diversity is measured as the average pairwise differences per site, $\pi$, in the sample, and is normalized using divergence between humans and outgroup species (see Materials and Methods). The purpose of this normalization is to account for the possibility that different parts of the genome may have different mutation rates. The mutation rates could systematically differ across chromosomal types because the different chromosomes spend different amounts of time in the male and female germlines and the male germline has a higher mutation rate than the female germline [30]. Because the low diversity on the Y chromosome persists after this normalization, it cannot be explained by a correspondingly low mutation rate on the Y chromosome (Table S2; Figure S1). Further, the highly repetitive ampliconic regions of the Y were not assembled by Complete Genomics, and so are not analyzed here (Materials and Methods). Diversity on the Y chromosome is likely not being under-estimated due to the inability to call variants in haploid regions of the genome because diversity on the X measured in females, where the X is diploid, is nearly identical to diversity on the X measured in males, where the X is haploid (Figure S2). The pattern of reduced diversity on the Y chromosome is observed in both Africans and Europeans, suggesting that the effect is not population-specific, and holds regardless of whether the neutral sequence analyzed is near or far from genes (Table 1). Previous analyses of portions of the Y reported low Y diversity [12-16], but measuring divergence-normalized $\pi$ per site at 0.0018 for Africans and 0.0024 for Europeans, we observe that chromosome-wide Y diversity is an order of magnitude lower than the equilibrium neutral expectation of one-quarter the autosomal level of diversity (Figure 1). Conversely, mitochondrial diversity is not reduced compared to expectations under neutrality (Figure 1). Additionally, our estimates of diversity on the X chromosome are consistent with previous estimates from Africans [5,17] and Europeans [3,5]. These trends held for all populations sampled in the public Complete Genomics data (Figure S3).

In contrast to diversity in other genomic regions, we observe that diversity is lower on the Y chromosome for the African populations in our sample than for the European populations in our sample (Table 1). Previous studies of Y chromosome diversity have also suggested that the difference in diversity on the Y is small between Africans and Europeans [31,32], or that it may, as we observe, be higher in Europeans than some African populations [15,33]. For example, haplotype diversity was found to be higher across Europeans than Africans (0.852 versus 0.841) [33]. Similarly, when the African populations are broken down into Sub-Saharan Africans versus North Africans (the Complete Genomics samples are Western/Northern Africans), European diversity falls in between these two, with European diversity on the Y chromosome actually higher than diversity in North Africans [33]. Other studies have observed slightly higher diversity in Africans than Europeans, but include a much more diverse group of Africans. For example, variation on the Y chromosome has been reported previously to be only slightly higher on the Y for African versus Non-African populations, even though the population of Africans is much more diverse (including Bakola from Cameroon, Dogon from Mali, Bantu from South Africa and Khoisan from Namibia and South Africa) [32] than the population we analyze. The uncorrected levels of diversity reported here for the Y chromosome (Table S2), differ from some previous studies [15,31,34], but are not directly comparable to these studies because: 1) they were based on genetic markers that were chosen specifically because they have high mutation rates [15,31,34]; and, 2) the populations are different than the ones available for this study [34]. The absolute number of SNPs identified here is not reduced relative to other sequencing platforms [35]. In fact, overall diversity is similarly observed to be low on the Y using this other technology, but a larger TMRCA is estimated [35], perhaps because the Y seems to harbor pockets of hidden diversity [36].

We next consider several possible models that could explain this unexpectedly low amount of diversity found on the Y chromosome relative to other genomic regions. Such models include differences in the variance in reproductive success between males and females, purifying selection on the Y chromosome, and positive selection on the Y chromosome.

**Variance in reproductive success**

In principle, a greater variance in male reproductive success than female reproductive success ($N_f > N_m$) could result in a lower than expected effective population size of the Y chromosome. In fact, previous studies have suggested that increased variance in offspring number has reduced the effective population size in human males versus females and might explain the reduced variability on the paternally inherited Y chromosome [4,13]. To test the hypothesis that sex-biased demography explains the decreased Y chromosome diversity, we modeled increasingly skewed sex ratios using coalescent simulations, taking into account the complex demography of the populations analyzed here (Figure 1; Table S3; Methods). We use the case where $N_m = N_f$ as the null model. As expected, decreases in the male effective population sizes ($N_m/N_f<1$) decrease expected Y diversity. However, we find that the reduction in the male effective population size required to explain the observed Y chromosome data, predicts levels of normalized autosomal, X and mtDNA diversity that are not consistent with the data in these markers (Figure 1; Table S3). This effect can also be illustrated by considering ratios of normalized diversity in each type of marker relative to autosomes. A skew in the sex ratio large enough to explain the observed reduction in Y/autosome diversity would also cause increases in X/autosome and mtDNA/autosome diversity that are incompatible with observations (Figure 1; Table S4). Thus, by analyzing all classes of genomic sequences, we are able to reject extreme sex-biased processes as the sole explanation for patterns of low observed Y variability.

**Purifying selection**

Natural selection has also been suggested to play a large role in reducing diversity on the Y chromosome [12,14-16], and works within the context of the demographic history of the populations. Purifying selection can reduce genetic variation at linked neutral sites via a process called background selection, which has received extensive theoretical treatment in the literature [21,22,26,37-41]. Purifying selection has already been documented for the mtDNA [42]. Due to the lack of homologous recombination throughout most of the Y chromosome, background selection is expected to have a particularly strong effect, severely reducing diversity on the Y chromosome. Two factors determine the overall effect of background selection on reducing neutral diversity in non-recombining regions: 1) The strength of selection, and 2) the number of sites subject to selection. At approximately 60 million base pairs, there are orders of magnitude more sites that may be subject to selection on the human Y chromosome than on the mtDNA. Selection may actually be quite weak on individual mutations that occur on the Y chromosome, but in the absence of recombination, if many sites are possible targets of this weak selection, this can lead to a strong reduction in diversity among Y chromosomes.

Here, we performed forward simulations with purifying selection to assess whether background selection could reduce diversity at neutral sites on the Y chromosome to the levels observed in our data. We study purifying selection under different assumptions of the variance in male reproductive success. We chose to use forward simulations, rather than using standard analytical background selection models, which assume the effect of background selection is a simple reduction in effective population size, for several reasons. First, the standard formulas were derived for equilibrium demographic models, but human populations have a more complex demographic history with unknown effects on the process of background selection. Second, many mutations have been shown to be weakly deleterious and may persist in the population due to genetic drift [37,38]. The standard theory does not allow for this. Finally, simulations studies suggest that the standard theory can over-predict the reduction in genetic diversity due to background selection if there are many weakly selected linked mutations [26]. The forward simulations that we conducted address all of these concerns.

We first evaluated whether purifying selection acting only on new nonsynonymous mutations in the coding regions of the Y chromosome could reduce levels of genetic diversity at linked neutral sites to the levels detected in our observed Y chromosome data. To do this, we performed forward simulations using realistic demographic models for the populations where only new nonsynonymous mutations were subjected to purifying selection (see Methods). We find that models of selection acting only on coding sites cannot sufficiently reduce expected diversity at linked neutral sites through background selection on the Y chromosome. Under the assumption of equal sex ratios, regardless of the mean selection coefficient used, all models result in levels of diversity at linked neutral sites that are significantly higher than the observed values for both Africans ($P<0.001$) and Europeans ($P<0.025$, Figure S4).

In principle, models with a larger female effective population size could explain the low diversity observed on the Y chromosome. However, we have demonstrated that such models cannot match the levels of genetic diversity observed on the X chromosome, mtDNA, and Y chromosome together. However, sex-biased demography along with purifying selection acting on new nonsynonymous mutations in the coding regions of the Y chromosome could reduce levels of diversity at linked neutral sites. To evaluate the joint effects of sex-biased demography and purifying selection, we used levels of putatively neutral diversity (*i.e.*, diversity far from genes) on the X chromosome and the autosomes to estimate the degree of sex-biased demography for the populations in our study (Table 2). We find that $N_m/N_f = 0.335$ in the African population which is concordant with estimates from previous studies [4,20,35]. Under an assumption of an extremely reduced male effective population size, relative to females ($N_m/N_f = 0.335$) which matches patterns of diversity on the X chromosome, predicted diversity at linked neutral sites, from models including purifying selection only on nonsynonymous mutations, is still significantly higher than the observations in Africans (P<0.001, Figure S4). In Europeans, we estimate that that $N_m/N_f = 1$ (Table 2). These results hold for a wide range of the mean strength of selection (Methods; Figure S5).

**Estimating the number of sites under purifying selection on the Y chromosome**

Given its unique structure, it is possible that purifying selection acts on more than just the nonsynonymous sites on the Y chromosome. Specifically, in addition to the approximately 100,000 single copy coding sites (predicted from annotated coding genes [43]; Methods), the Y also contains 5.7Mb of highly repetitive ampliconic regions, composed of long palindrome "arms", each with nearly-identical sequences [27,28]. Genes in these ampliconic regions are expressed exclusively in the testis [27,28], and so may be under selection related to male fertility. Further, it has been hypothesized that, in the absence of homologous recombination with the X, intra-chromosome pairing and the resulting gene conversion between palindrome arms may reduce the mutational load on the Y, and so these palindromes themselves, as a means of allowing intra-chromosome recombination, may be subjects of selection [27-29].

Thus, we developed a novel approximate likelihood approach to estimate the number of sites affected by purifying selection (*L*) required to reduce diversity at linked neutral sites to the low values observed on the Y (Methods). Simulations show that our method can accurately estimate *L* (Methods; Table S5). Assuming an equal sex ratio, the maximum likelihood estimate of the number of sites subjected to purifying selection on the Y is as much as 30 fold higher than the number of coding sites, for both Africans and Europeans (Figure 2). Relaxing the assumption of an equal sex ratio to allow many fewer males relative to females (to the ratio of the number of males to the number of females that fit neutral diversity on the X and autosomes, $N_m/N_f = 0.335$ [4,20]), and to an extreme bias in male reproductive success of $N_m/N_f = 0.1$, slightly decreases the estimates of the number of sites directly affected by purifying selection. However, the estimate from the African sample is still significantly greater than the number of coding sites. Our results strongly support the hypothesis that at least some of the ampliconic regions evolve under the direct effects of purifying selection, where new mutations in these regions are deleterious.

The above estimates assume that the selection coefficients of the deleterious mutations on the Y chromosome are the same as those estimated from nonsynonymous mutations on the autosomes, with appropriate re-scaling to account for the differences in Ne and ploidy on the autosomes and the Y chromosome (see Methods). However, it is possible that the strength of selection acting on noncoding mutations on the Y chromosome could be different than that acting on nonsynonymous mutations on the autosomes. It is unclear whether this difference in the strength of selection could bias our estimates of the number of sites directly under selection. To address this concern, we extended our approach to jointly estimate the number of sites directly affected by purifying selection (L) as well as the mean strength of selection (see Methods). Even when considering a range of different strengths of selection, we find that the estimates of the number of sites to be directly under the effect of purifying selection are largely insensitive to the mean strength of selection, and are still more than the number of X-degenerate coding sites (Figures S5 and S6). This suggests that content recruited to the Y chromosome after X-Y recombination was suppressed, including the high-copy-number ampliconic regions, as well as any transcription factor binding sites, may be subject to purifying selection that, due to the lack of homologous recombination, acts to reduce diversity on the human Y chromosome.

We found that a population expansion model matched the average observed levels of autosomal, X and mtDNA polymorphism in the African populations, and a bottleneck model matched the observed levels of polymorphism in the European population (Figure 1, Tables S4, S5 and S7). Several publications have documented various signatures of background selection throughout the genome [17,44-47]. If background selection had reduced average levels of diversity across the genome (previous work suggests around a 6% reduction in diversity [24]), this would mean that the demographic parameters that fit the data were not truly reflective of population history, but instead reflected both population history and background selection. Thus, even if background selection is operating on the putatively neutral genomic regions we analyze here, the reduction in diversity on the Y chromosome is still too extreme to be consistent with that level of background selection. Rather, additional background selection, as we have modeled here, would be required.

**Positive selection**

Although models of purifying selection are consistent with the low observed diversity, it is also possible that positive natural selection may also be driving low diversity on the human Y via selective sweeps [48,49], when neutral variation is removed due to the fixation of an advantageous mutation. Although it can be difficult to distinguish between genetic signals of background selection versus positive selection with few nucleotide polymorphisms, as is the case with the Y chromosome, we analyzed the data using two additional measures. First, we computed the folded site frequency spectrum for Y chromosome SNPs across all unrelated Y chromosomes in the Complete Genomics dataset (Figure S8). The abundance of low frequency SNPs is consistent with both positive selection and purifying selection (Figure S8), and the low overall number of SNPs makes further distinctions between the two models difficult. Second, we built a neighbor-joining tree for all unrelated Y haplotypes in the Complete Genomics dataset using phylip [50], then branch lengths were computed using a molecular clock in paml [51]. There is not an overarching star phylogeny, which would be indicative of a single selective sweep (Figure S9). While we cannot rule out such a scenario directly, we note that previous studies also found little or no evidence of selective sweeps [52] or gene-specific positive selection [53,54] on the Y chromosome. However, one might conceive of a complex evolutionary history involving several instances of positive selection along different Y lineages that could result in the observed haplotype topology. Given recent findings of pockets of Y haplotype diversity, it is possible that recurrent positive selection may contribute to reduced Y diversity [36].

**Conclusions**

We observe that diversity across the entire human Y chromosome is extremely low. We find that neutral models with sex-biased demography may contribute to low Y diversity. However, models of extreme differences in reproductive success between males and females are insufficient as the sole explanation for patterns of genome-wide diversity. Alternatively, then, natural selection appears to be acting to reduce diversity on the Y. We show that models of purifying selection affecting Y chromosome diversity are consistent with low observed diversity, if purifying selection acts on more than the few coding regions left on the Y chromosome. Thus, our results suggest that selection may also act on the highly repetitive ampliconic regions, and support arguments for the functional importance of these regions [29]. Further strong purifying selection acting on the human Y is consistent with reports of the conservation of both the number and the type of functional coding genes on the Y chromosome in humans [12] and across primates [55,56]. It is also possible that positive selection has been acting to reduce diversity on the Y chromosome, but this explanation would require multiple independent selective sweeps across populations.

Although positive selection is expected to confound evolutionary relationships, if purifying selection is the dominant force on the Y chromosome, the topology of the tree should remain intact, but the coalescent times are expected to be reduced. This means that the Y chromosome, keeping in mind that it is a single marker without recombination, may actually provide a more useful marker for inferring phylogeographic patterns than other markers. Indeed, recent resequencing efforts of the Y chromosome identified a single mutation that resolves a previously unresolved trifurcation of lineages, and reports monophyletic groupings of Y chromosomes from distinct populations [35]. While it a combination of factors influence genome-wide estimates of diversity, and variance in male reproductive success still affects patterns of autosomal, X, Y and mtDNA diversity, selection clearly affects levels of diversity on the Y, and so should be considered when drawing conclusions regarding demography and population history based on patterns of Y-linked markers.

**Materials and Methods**

**Genomic data analysis**

We analyzed unrelated, high quality, publicly available whole genomes generated by Complete Genomics assembly software version 2.0.0 [57] (Table S1). Next generation sequence data often suffer from sequence errors, assembly errors and missing information, and non-reference alleles will be less likely to be mapped [58]. However, the Complete Genomics dataset overcomes many of these errors by using very high coverage (> 30X [57]). Additionally, to be conservative, we only consider sites with data called in all individuals in each population. We removed putatively functional and difficult to assemble regions including: RefSeq known genes, CpG islands, simple repeats, repetitive elements (RepeatMasker), centromeres, and telomeres, downloaded from the UCSC Genome browser [43], and filtered using Galaxy [59]. We also excluded the hypervariable regions on the mtDNA [60], which might inflate estimates of mitochondrial diversity, and analyzed only the X-degenerate regions of the human Y [27], because diversity might be reduced in the pseudoautosomal or ampliconic regions. Divergence was computed from number of nucleotide differences per site between pairwise human and chimpanzee reference sequence alignments for autosomes, X, and mtDNA downloaded from the UCSC genome browser [43], and for the Y from ref [28]. The total number of SNPs called on the Y chromosome in the Complete Genomics dataset does not appear to be lower than other chromosome-wide assessments of Y variation. Of the SNPs across 16 individuals that overlap between the 1000 genomes (252 SNPs) and Complete genomics dataset (6236), there are only 12 sites called in the 1000 genomes dataset that are not called in the Complete Genomics dataset; all of these are singletons, and many have missing data across several individuals (Table S7). Further, the geographic distribution of Y chromosome sampled for the Complete Genomics dataset does not appear to be wider for the European versus the African populations [61]. The per generation per site mutation rates estimated from human-chimpanzee alignments, assuming a divergence time of 6 million years and 20 years per generation, are $2.11 \times 10^{-08}$ for the autosomes, $1.65 \times 10^{-08}$ for chromosome X, and $3.42 \times 10^{-08}$ for chromosome Y. For mtDNA we use the mutation rate reported of $1.7 \times 10^{-08}$ for the mtDNA [62]. The recombination rates used were 1cM/Mb and (2/3)*(1cM/Mb), for the autosomes and X, respectively. Diversity is measured using, π, the average number of nucleotide differences per site between all pair of sequences. For the inference of the number of sites under selection, we summarize the genetic variation data by $S$, the number of segregating sites, because the distribution of $S$, conditional on the underlying genealogy, is known (Poisson, see below). We do not directly analyze the ampliconic regions, as they were not assembled in the Complete Genomics data.

All estimates of diversity, and human-chimpanzee divergence used for normalization are reported in Table S2. Human-orangutan estimates of divergence could not be used because no whole Y chromosome sequence currently exists for orangutan. Although the Y chromosome sequence was recently published for the rhesus macaque, the sequence has diverged and degraded so much between human and macaque that very little of the noncoding regions are alignable [55], preventing us from reliably correcting for divergence across all chromosome types using human-macaque divergence.

**Modeling male and female effective population sizes**

Population genetics parameters used in coalescent [63] and forward simulations [64] for Europeans and Africans are similar to previously published estimates [65,66]. We use a simple model of drift, which assumes purely random (Poisson) variation in offspring numbers for both males and females, and non-overlapping generations. For Africans, the neutral model is of an expansion from 10,000 to 20,000 individuals 4,000 generations ago. For Europeans the neutral model is of a bottleneck from 10,000 to 1,000 individuals 1,500 generations ago, followed by an expansion to 10,000 individuals 1,100 generations ago (Table S6).

Neutral expectations under equal and skewed sex ratios were modeled using coalescent simulations implemented in ms [63], assuming the population-specific demographic models described above, and allowing for recombination on the autosomes and X chromosome, but not the Y or mtDNA.

The effective population sizes for each chromosome type ($N_{auto}$, $N_{chrX}$, $N_{chrY}$, and $N_{mtDNA}$), for given male and female effective population sizes ($N_m$ and $N_f$) are (see *e.g.*, ref [67]):

$$N_{auto} = 4N_m N_f / (N_m + N_f)$$
$$N_{chrX} = 9N_m N_f / (4N_m + 2N_f)$$
$$N_{chrY} = N_m / 2$$
$$N_{mtDNA} = N_f / 2$$

For a fixed ratio and males to females ($R = N_m/N_f$), and fixed total effective population size ($N_{auto}$), we then calculate the male and female effective population sizes as:

$$N_f = N_{auto} (1 + R) / (4R)$$
$$N_m = N_f R,$$

Using these equations we can use standard neutral coalescent simulations implemented in ms to simulate data for the four chromosome types, while varying $R$, but keeping $N_{auto}$ constant. We keep $N_{auto}$ constant to mimic the real data, as the demographic parameters were originally estimated from autosomal markers. Further details about the values used for simulations can be found in Table S8. Complete commands for ms simulations are given in Note S1.

**Modeling purifying selection**

We modeled purifying selection using forward simulations implemented in SFS_CODE [64]. The exact commands used in the SFS_CODE simulations are given in Note S1. Similar to the coalescent simulations, we modeled the African and European populations separately, using the population-specific demographic models described above, the Y chromosome per generation per base pair mutation rate, and sampling 8 chromosomes per simulation to match the sample size of our observed data. However, unlike ms, which scales parameters by the current population size and moves backward in time, SFS_CODE starts with the ancestral number of chromosomes and simulates a haploid population forward in time. Thus, when rescaling the effective population size from the autosomal estimates, for SFS_CODE we used the same diploid autosomal ancestral effective population size for both populations ($N$ =10,000). The Y chromosome effective size was then found using the same process described above for the neutral coalescent simulations.

**Evaluating purifying selection on coding sites**

To investigate purifying selection acting only on new nonsynonymous mutations, we simulated 60,041 nonsynonymous sites (90,062 coding sites are estimated from the union of all exons from X-degenerate, non-pseudoautosomal genes on the Y chromosome [43]) at which new mutations are expected to be subject to purifying selection. To assess the effect of background selection, each simulation also contained 500kb of linked neutral sequence from which we calculated diversity.

The effect of background selection is a function of the distribution of selection coefficients for new, deleterious mutations, and can be modeled by varying the mutation rate, the number of sites affected by selection (*L*), and the selection coefficient acting on new mutations (*s*) [21]. When evaluating models with different strengths of purifying selection, we assumed that selection coefficients for the nonsynonymous sites were drawn from a gamma distribution. Previous studies found this distribution to fit the observed autosomal frequency spectrum well [37,38,68,69], and there is little reason to believe that the shape of the gamma distribution varies across chromosomes. However, although the X- and Y-linked genes are often highly diverged in sequence and function, the remaining X-degenerate Y-linked genes are likely highly constrained in order to have survived on the Y [70]. Thus, it may not be precise to assume X-degenerate Y-linked genes evolve under similar selective constraints as autosomal genes. To address this, we investigate a wide range of scale parameters of the gamma distribution. For a fixed value of the shape parameter of the gamma distribution, the mean strength of selection can be changed by modifying the scale parameter of the gamma distribution. Thus, we fixed the shape parameter to 0.184 (as estimated by refs [37,69]) and performed simulations using mean selection coefficients ranging from 0.0001 to 0.09 (Figure S4).

We ran 1,000 replicates for each set of selection parameters in each population. For each replicate we calculated $\pi^*$, the simulated per site nucleotide diversity (average number of pairwise differences) normalized by the per site human-chimp divergence (0.02051; Table S1). The similarly calculated observed Y diversity is denoted $\pi_{obs}$. For each set of parameter values we then calculated $P_1$, the proportion of simulation replicates with $\pi^* > \pi_{obs}$ was used to calculate a 2-sided *P*-value by $P_2 = 1 - 2 \times |P_2 - 0.5|$. Models with could not be rejected and were considered to fit the observed data.

**Estimating the number of sites under purifying selection**

To estimate the number of sites directly affected by purifying selection on the Y chromosome (defined as *L*) from looking at the levels of diversity at linked neutral sites, we developed a novel approximate likelihood approach [65,71,72] using the observed number of segregating sites, $S_{obs}$, in neutral regions, as a summary statistic. We then define the likelihood function for *L* in a neutral region as:

$$\Pr(S_{obs} = S | L, \Theta) = \int_0^\infty \Pr(S_{obs} = S | T) p_T(T | L, \Theta) dT.$$

where is the number of segregating sites in neutral regions of the observed data, is the sum of all the branch lengths of the genealogy in units of generations, and refers to all of the other fixed parameters in the model (*e.g.*, the demographic history and distribution of selection coefficients). Under the infinite sites model, the conditional distribution of $S_{obs}$ given *T* is Poisson (see e.g., [73]):

$$\Pr(S_{obs} = S | T) = \frac{(\mu T)^S}{S!} e^{-\mu T}$$

where μ is the neutral mutation rate per generation over the entire region. This relationship holds even if the underlying genealogy has been affected by natural selection or other non-stationary demographic processes, as long as the individual mutations being analyzed are neutral. Then, the number of sites affected by purifying selection, *L*, enters the likelihood function by the effect that selection has on the genealogy. $p_T(T | L, \Theta)$ is the distribution (density) of the sum of the branch lengths over the entire genealogy under the particular model of demography and selection, with $L$ sites directly affected by purifying selection. This distribution is difficult to calculate directly, and in general, the integral given above cannot be solved analytically. However, it could be approximated using simulation approaches that keep track of the genealogy as part of a forward simulation method [74]. If we could simulate $p_T(T|L,\Theta)$ from, then the distribution of could be approximated as the sum:

$$\frac{1}{k}\sum_{i=1}^{k}\frac{(\mu T_i^*)^{S_{obs}}}{S_{obs}!}e^{-\mu T_i^*}$$

However, even such an approach is cumbersome and slow because of the overhead involved in keeping track of a genealogy in simulations with multiple loci under selection. We instead employ an approximate approach using forward-simulations implemented in SFS_CODE [64]. For a simulation replicate producing variable sites, and with a simulated value of equal to $T^*$,

$$\lim_{\mu\to\infty}\Pr\left(\left|\frac{S^*}{\mu}-T^*\right|>\varepsilon\right)=0.$$

Therefore, a simulation consistent estimator of can be obtained from the number of segregating sites in a simulated sample. In other words, if we simulate enough sites in each replicate, the total tree length can be approximated using the number of segregating sites (Table S5; Figure S5). The aforementioned integral in the likelihood function can therefore be approximated stochastically by simulating data sets using SFS_CODE, with $S_i^*$, = 1, 2,…$k$, segregating sites, and each with a neutral mutation rate of $\mu_{sim}$, and then evaluating,

$$\frac{1}{k}\sum_{i=1}^{k}\frac{((\mu/\mu_{sim})S_i^*)^{S_{obs}}}{S_{obs}!}e^{-(\mu/\mu_{sim})S_i^*},$$

as an estimator of the likelihood function for $L$ based on $S_{obs}$. The number of neutral base pairs on the Y chromosome with sufficient sequencing data was 7,758,906 and 7,974,045 bp for the African and European populations respectively. Assuming a neutral mutation rate of $3.42 \times 10^{-08}$ per base pair per generation, $\mu = 0.265$ for the African population and $\mu = 0.273$ for the European population. However, forward simulations of >7Mb of sequence are extremely time consuming. Thus, for computational efficiency, we simulated 500kb of neutral sequence, giving $\mu_{sim} = 0.0171$. We accounted for the fact that we simulated fewer neutral sites than in the actual data by including the ratio of the two per region mutation rates ($\mu/\mu_{sim}$), in our likelihood function represented above. We chose to simulate 500kb of neutral sequence because a region of that size is small enough to be computationally efficient while still allowing an accurate approximation of $T$ (Figure S5). Using this method we optimized the likelihood function over a grid of values for $L$ ranging from below the number of coding sites, 50kb, to more than the number of ampliconic regions, 6Mb.

The population scaled selection coefficient ($Ns$) acting on a particular deleterious mutation was drawn from a gamma distribution, with the parameters estimated in Boyko *et al.* [37], including the same shape parameter (0.184) used above. However, because the Boyko *et al.* [37] model was developed for the autosomes, and assumes semi-dominant effects, we rescaled the mean strength of selection for a haploid model to represent Y evolution. The scale parameter of the Boyko *et al.* model (8200) was divided by the ratio of the number of chromosomes used in the original model (51272) to the number of Y chromosomes used in our simulations (5000), then multiplied by 2 because the original model described the fitness of a mutation in the heterozygous state, and all mutations on the Y chromosome will immediately be exposed to selection. Thus, our model used the resulting scale parameter (1600).

We also jointly estimated the number of sites directly under selection (*L*) and the mean strength of selection by looking at neutral diversity levels on the Y chromosome. We employed an approximate likelihood approach similar to that described above. However, here we investigated a two-dimensional grid of different values for *L* and a grid of different scale parameters for the gamma distribution of selective effects. Because we kept the shape parameter fixed at 0.184, changing the scale parameter changed the mean strength of selection. We found that our estimates of *L* were largely insensitive to the mean strength of selection. The profile likelihood curve shown in Figure S7 is remarkably similar to the likelihood curve shown in Figure S6, when the mean strength of selection was held constant.

Asymptotic approximate 95% confidence intervals included all points in the log-likelihood curve that fell within 1.92 log-likelihood units from the MLE (Note S1; Figure S6). Linear interpolation was used to find the appropriate cutoff in between grid points. SFS_CODE commands used for this section are given in Note S1.

**Performance of the approximate likelihood approach on simulated data**

We performed simulations to evaluate the performance of our approximate likelihood approach to estimate *L* by simulating 1,000 Y chromosome datasets using SFS_CODE under models of African and European demographic history. No recombination was allowed on the Y chromosome. Each simulation replicate, or simulated dataset, included 7.5 Mb of neutral sequence (equivalent to the size of our observed data) linked to 2Mb of sites (*i.e.*, *L* = 2Mb) where new mutations were subjected to purifying selection (with selection coefficients drawn from the gamma distribution as discussed in Methods). For each simulated region, the approximate likelihood approach was used to estimate *L* based on the number of segregating sites within the neutral region. The distribution of selection coefficients used in the inference procedure was the same distribution used to simulate the data. The mean and median of the maximum likelihood estimates (MLEs) as well as the coverage properties of the asymptotic 95% confidence intervals (CIs) are shown in Table S5. The asymptotic 95% CIs contain the true value of *L* 96.6% of the time for the African simulations and 98.3% of time for the European simulations (rather than 95% of the time), suggesting that they are slightly conservative.

**Testing models of sex-biased demography and purifying selection**

We repeated our analyses of whether purifying selection on coding sites can explain the low diversity on the Y chromosome and our estimation of the number of sites affected by purifying selection taking into account unequal male and female population sizes. We also evaluated whether the low diversity on the Y chromosome could be accounted for by purifying selection combined with unequal male and female population sizes. In particular, Hammer *et al.* [17] and Lohmueller *et al.* [20] estimate that there were roughly 2.63 females reproducing for each male that reproduces. In other words, $N_m=0.38N_f$. Additionally, we performed our own estimate of $N_m/N_f$ from the levels of diversity at putatively neutral sites (those >100kb from genes) on the X chromosome and the autosomes and estimate $N_m=0.3352N$ (Table 2). We have shown (Figure 1) that demographic models with an autosomal ancestral effective population size of roughly 10,000 individuals fit the autosomal levels of diversity reasonably well (Table S3). We compute the effective population size of males under a skewed sex ratio by inputting the previously observed $N_m/N_f$ ratio of 0.3352, and the autosomal size of 10,000 individual, in the equation [67]:

$$N_A = \frac{4N_m N_f}{N_m + N_f}, N_m = 3338.$$

We then repeated the forward simulations and analyses described above using this value for $N_m$.

**Acknowledgements**

We would like to thank Brian Charlesworth and our anonymous reviewers for their careful and thoughtful comments that have improved this study.

**Tables**

**Table 1. Observed and mean modeled estimates of neutral diversity.** For Africans, the neutral model is of an expansion from 10,000 individuals to 20,000 individuals 4,000 generations ago. For Europeans the neutral model includes a bottleneck from 10,000 individuals to 1000 individuals 1,500 generations ago, followed by an expansion to 10,000 individuals 1,100 generations ago. Expected values assume the null model of equal numbers of breeding males and females ($N_m/N_f = 1$). Diversity is normalized by human-chimpanzee divergence, to correct for different mutation rates for each class of sequence. Diversity is additionally calculated far from genes (100kb away), except for on the mtDNA, where doing so would eliminate all sequence from analysis.

| | Diversity | | | | | |
|---|---|---|---|---|---|---|
| | African | | | European | | |
| | Observed | | | Observed | | |
| chr | $\pi$ | $\pi_{Far}$ | Modeled | $\pi$ | $\pi_{Far}$ | Modeled |
| A | 0.0739 | 0.0804 | 0.0733 | 0.0563 | 0.0615 | 0.0561 |
| X | 0.0601 | 0.0723 | 0.0565 | 0.0365 | 0.0459 | 0.0403 |
| Y | 0.0018 | 0.0019 | 0.0223 | 0.0024 | 0.0027 | 0.0100 |
| M | 0.0235 | 0.0235* | 0.0222 | 0.0147 | 0.0147 * | 0.0101 |

* If the "far from genes" filter were applied to the mtDNA, there would be no sequence left to analyze

**Table 2: Estimates of $N_m/N_f$ using X and autosomal genetic diversity far from genes.**

| Population | $\pi_X$ | $\pi_A$ | $\pi_X/\pi_A$ | $N_m/N_f$ |
|---|---|---|---|---|
| Africa | 0.0723 | 0.0804 | 0.8993 | 0.3352 |
| Europe | 0.0459 | 0.0615 | 0.7463 | 1.0299 |

**Figures Legends**

**Figure 1. Observed and expected ratios of normalized X/Autosome, Y/Autosome, and mtDNA/Autosome nucleotide diversities.** The expected values under an equal male/female ratio for X/Autosome ratio (0.75) and for Y/Autosome and mtDNA/Autosome (0.25) are plotted for reference. Twice the standard error is plotted for each model, computed from the ratios of 10,000 replicates per chromosome comparison. Expected values were computed from simulations using different demographic histories for Africans and Europeans (Tables 1, S4 and S5), first assuming equal numbers of males and females ($N_m/N_f = 1$), then successively skewing the effective number of males relative to females in each population (e.g. $N_m/N_f = 0.75$ implies three males for every four females in the population). All chromosomes were normalized for chromosome-specific mutation rates using divergence from chimpanzee.

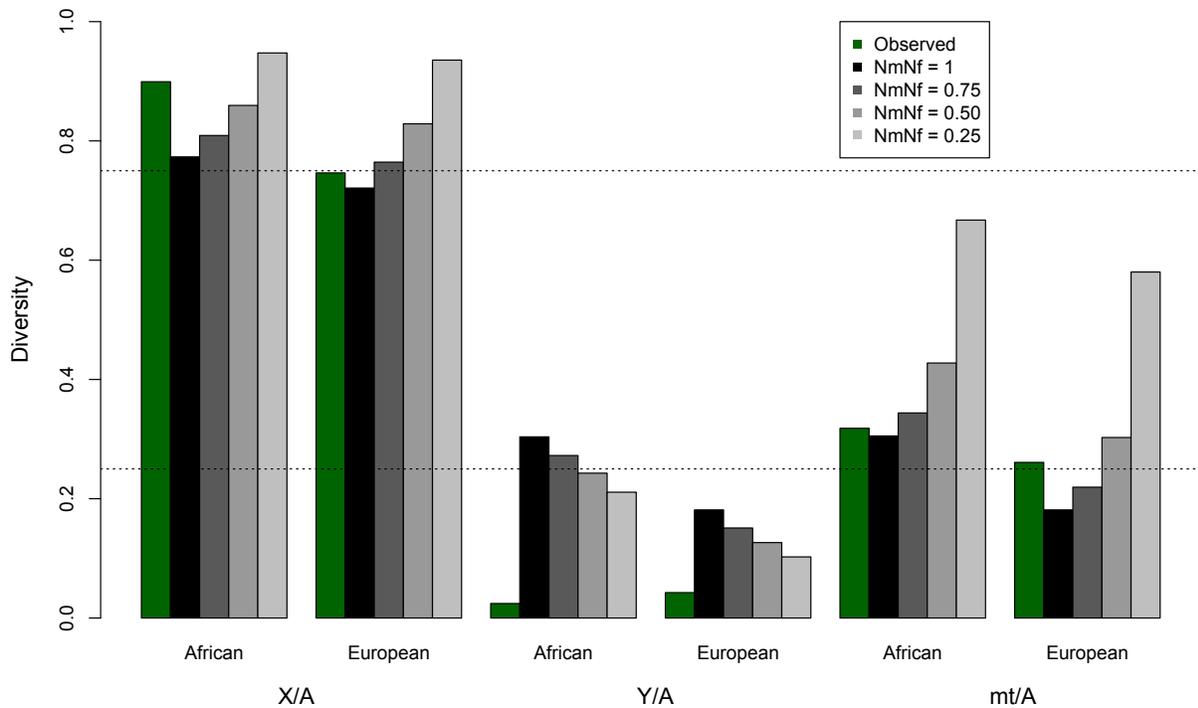

**Figure 2. Estimates of the number of sites affected by purifying selection ($L$) on chromosome Y.** The maximum likelihood estimates (MLEs) and 95% confidence intervals of the number of sites affected by purifying selection on the Y chromosome are plotted for Africans (red) and Europeans (blue). Assuming no sex-biased demography, the MLE for Africans is 5 Mb (95% CI: 1.36-6 Mb) and for Europeans it is 3 Mb (95% CI: 0.798-6 Mb). Estimates were made assuming an equal sex ratio ($N_m/N_f = 1$), and assuming a highly skewed sex ratio ($N_m/N_f = 0.38$). Assuming this sex-biased demography, the MLE for Africans is 5 Mb (95% CI: 1.85-6 Mb), and for Europeans it is 2 Mb (95% CI: 0.18-4.2 Mb). The number of ampliconic and coding sites on the Y chromosome are plotted in horizontal dotted lines.

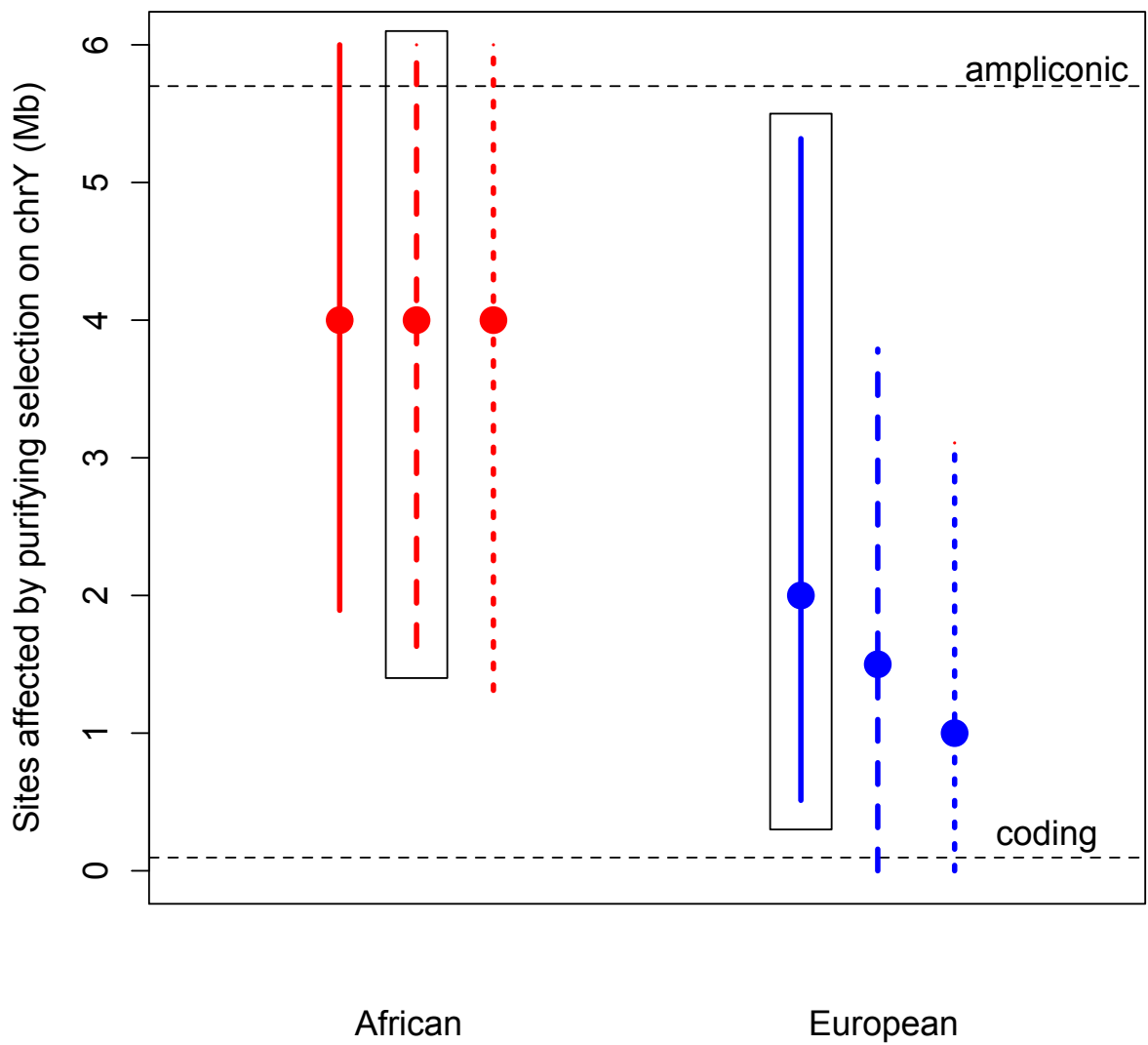

**Supplementary Information Legends**

**Note S1. Simulation command line codes.** Command lines for coalescent and forward simulations are described in detail.

*Simulation command line codes*

Neutral expectations under equal and skewed sex ratios were modeled using coalescent simulations implemented in ms [63], assuming the population-specific demographic models described above, and allowing for recombination on the autosomes and X chromosome, but not the Y or mtDNA (Methods; Table S7). The ms commands used take the forms below:

For Africans:
    A & X ./ms *$indv $repl* -t *$theta* -r *$rho $sites* -eN *$time1 $size1*
    Y & M ./ms *$indv $repl* -t *$theta* -r *$sites* -eN *$time1 $size1*

For Europeans:
    A & X ./ms *$indv $repl* -t *$theta* -r *$rho $sites* -eN *$time1 $size1* -eN *$time2 $size2*
    Y & M ./ms *$indv $repl* -t *$theta* -eN *$time1 $size1* -eN *$time2 $size2*

Where:

    $indv = 16 for autosomes; 8 for X, Y and mtDNA;
    $repl = 10000;
    $theta = 4×$N×$u×$sites;
    $rho = 4×$N×$r×$sites;
    $time = African:    $time1 = 4000/(4*$N);
            European:    $time1 = 1100/(4*$N);
                           $time2 = 1500/(4*$N);

    $size = African:    $size1 = 0.5;
            European:    $size1 = 0.01;
                           $size2 = 1;

We modeled purifying selection using forward simulations implemented in SFS_CODE [64]. To investigate purifying selection acting only on new nonsynonymous mutations, we simulated 60,041 nonsynonymous sites (2/3 the 90,062 coding sites estimated on the Y chromosome [43]) at which new mutations are expected to be subject to purifying selection. To assess the effect of background selection, each simulation also contained 500kb of linked neutral sequence from which we calculated diversity. The SFS_CODE commands for modeling purifying selection on the coding sites are:

For Africans:
    ./sfs_code 1 50 -r 0 -t 0.000170954 -P 1 -TE 0.8 -Td 0 2.0 -L 2 500000 60041 -a N -W L 1 2 0 1 1 0.184 $ln -n 8 -N 500 –A

For Europeans:

./sfs_code 1 125 -r 0 -t 0.000170954 -P 1 -TE 0.3 -Td 0 0.1 -Td 0.08 10 -L 2 500000 60041 -a N -W L 1 2 0 1 1 0.184 $ln -n 8 -N 500 -A

Where, in each simulation, $ln=0.184/($mean_s*5000).

To estimate the number of sites affected by purifying selection on the Y chromosome (defined as *L*), we employ an approximate likelihood approach [65,71,72] using the observed number of segregating sites as a summary statistic. The SFS_CODE commands used to estimate the number of sites under selection are:

For Africans:
./sfs_code 1 50 -r 0 -t 0.000170954 -P 1 -TE 0.8 -Td 0 2.0 -L 2 500000 $L -a N -W L 1 2 0 1 1 0.184 0.000625272494772688 -n 8 -N 500 -A

For Europeans:
./sfs_code 1 125 -r 0 -t 0.000170954 -P 1 -TE 0.3 -Td 0 0.1 -Td 0.08 10 -L 2 500000 $L -a N -W L 1 2 0 1 1 0.184 0.000625272494772688 -n 8 -N 500 -A

$L is the number of sites subjected to purifying selection.

**Supplementary Tables**

**Table S1**. **Complete Genomics unrelated male samples.** IDs, sex, population, ethnicity, and abbreviations are provided for each of the Complete Genomics samples used. We cross-checked each individual to make sure there were no previously unreported relationships between them that might confound analyses [75].

| ID | Sex | Population | Ethnicity | Abbreviation |
|---|---|---|---|---|
| NA18501 | Male | African | YORUBA | YRI |
| NA18504 | Male | African | YORUBA | YRI |
| NA19020 | Male | African | LUHYA | LWK |
| NA19025 | Male | African | LUHYA | LWK |
| NA19026 | Male | African | LUHYA | LWK |
| NA19239 | Male | African | YORUBA | YRI |
| NA21732 | Male | African | MAASAI | MKK |
| NA21737 | Male | African | MAASAI | MKK |
| NA06994 | Male | European | UTAH/MORMON | CEU |
| NA07357 | Male | European | UTAH/MORMON | CEU |
| NA10851 | Male | European | UTAH/MORMON | CEU |
| NA12889 | Male | European | UTAH/MORMON | CEU |
| NA12891 | Male | European | UTAH/MORMON | CEU |
| NA20509 | Male | European | TOSCANI (TUSCAN) | TSI |
| NA20510 | Male | European | TOSCANI (TUSCAN) | TSI |
| NA20511 | Male | European | TOSCANI (TUSCAN) | TSI |

**Table S2**. **Uncorrected diversity (π) within Africans and Europeans, and human divergence from chimpanzee. All values are per site.** Estimates of the mutation rate for the mtDNA are corrected for multiple substitutions using the Tamura-Nei model.

| Class | A | X | Y | mtDNA |
|---|---|---|---|---|
| Africa diversity | 0.00093621 | 0.00059546 | 0.00003603 | 0.00169340 |
| European diversity | 0.00071314 | 0.00036196 | 0.00004901 | 0.00105811 |
| Sites in all 8 Africans | 1238501316 | 56363307 | 7758906 | 3004 |
| Sites in all 8 Europeans | 1246322049 | 56573306 | 7974045 | 3037 |
| Divergence from chimpanzee | 0.012664107 | 0.009911943 | 0.020514487 | 0.0720 |

**Table S3. Observed and mean modeled estimates of neutral diversity under various estimates of the $N_m/N_f$ ratio.** For Africans, the neutral model is of an expansion from 10,000 individuals to 20,000 individuals 4,000 generations ago. For Europeans the neutral model includes a bottleneck from 10,000 individuals to 1,000 individuals 1,500 generations ago, followed by an expansion to 10,000 individuals 1,100 generations ago. Mean estimates from 10,000 simulations under various assumptions of the ratio of the effective number of males to the effective number of females ($N_m/N_f$) are shown for Autosomes (A), chromosome X, chromosome Y and mtDNA.

|          |       |          | $N_m/N_f$ |        |        |        |        |        |
|----------|-------|----------|--------|--------|--------|--------|--------|--------|
| Pop      | Chr   | *Observed* | 1      | 0.9    | 0.75   | 0.5    | 0.38   | 0.25   |
| African  | A     | *0.0739* | 0.0733 | 0.0730 | 0.0730 | 0.0731 | 0.0730 | 0.0730 |
|          | chrX  | *0.0601* | 0.0565 | 0.0571 | 0.0588 | 0.0626 | 0.0652 | 0.0689 |
|          | chrY  | *0.0018* | 0.0223 | 0.0213 | 0.0198 | 0.0177 | 0.0167 | 0.0153 |
|          | mtDNA | *0.0235* | 0.0222 | 0.0233 | 0.0250 | 0.0311 | 0.0364 | 0.0485 |
| European | Auto  | *0.0563* | 0.0563 | 0.0563 | 0.0563 | 0.0564 | 0.0564 | 0.0564 |
|          | chrX  | *0.0365* | 0.0402 | 0.0411 | 0.0426 | 0.0462 | 0.0487 | 0.0522 |
|          | chrY  | *0.0024* | 0.0101 | 0.0094 | 0.0084 | 0.0071 | 0.0064 | 0.0057 |
|          | mtDNA | *0.0147* | 0.0101 | 0.0109 | 0.0123 | 0.0169 | 0.0219 | 0.0324 |

**Table S4. Observed and mean modeled ratios of neutral diversity under various estimates of the $N_m/N_f$ ratio.** For Africans, the neutral model is of an expansion from 10,000 individuals to 20,000 individuals 4,000 generations ago. For Europeans the neutral model includes a bottleneck from 10,000 individuals to 1,000 individuals 1,500 generations ago, followed by an expansion to 10,000 individuals 1,100 generations ago. Mean estimates from 10,000 simulations under various assumptions of the ratio of the effective number of males to the effective number of females ($N_m/N_f$) are shown for Autosomes (A), chromosome X, chromosome Y and mtDNA.

| Pop | Ratio | *Observed* | $N_m/N_f$ | | | | | |
| --- | --- | --- | --- | --- | --- | --- | --- | --- |
| | | | 1 | 0.9 | 0.75 | 0.5 | 0.38 | 0.25 |
| African | X/A | *0.8133* | 0.7732 | 0.7854 | 0.8091 | 0.8595 | 0.8962 | 0.9475 |
| | Y/A | *0.0244* | 0.3038 | 0.2933 | 0.2725 | 0.2429 | 0.2291 | 0.2110 |
| | mt/A | 0.3181 | 0.3053 | 0.3209 | 0.3439 | 0.4276 | 0.5006 | 0.6673 |
| European | X/A | *0.6483* | 0.7210 | 0.7383 | 0.7644 | 0.8287 | 0.8720 | 0.9355 |
| | Y/A | *0.0426* | 0.1813 | 0.1681 | 0.1510 | 0.1265 | 0.1142 | 0.1025 |
| | mt/A | *0.2610* | 0.1813 | 0.1965 | 0.2194 | 0.3027 | 0.3922 | 0.5802 |

**Table S5. Performance of our approximate likelihood approach to estimate *L* on simulated data.** *L* represents the number of sites affected by purifying selection. We assessed the accuracy of estimates by first making 1000 test datasets with a known value of *L* ($L = 2$Mb). We then computed the maximum likelihood estimate (MLE) of the number of sites affected by selection using our approximate likelihood approach. The table shows the mean and medians of the MLEs over the 1000 test datasets for each model. We also recorded the percentage of asymptotic 95% confidence intervals that contained the true value of *L*. The results, summarized in the table below, show that our method can accurately estimate the number of sites affected by purifying selection.

|  | Mean MLE | Median MLE | % of 95% CIs that contain the true *L* |
|---|---|---|---|
| African | 2.23 Mb | 2.00 Mb | 96.6% |
| European | 2.24 Mb | 2.00 Mb | 98.3% |

**Table S6. European observed and mean modeled estimates of diversity for various intensities of the population bottleneck.** The model is of neutral evolution with a bottleneck from 1500 generations ago to 1100 generations ago, from an ancestral size of 10,000 individuals, and a contemporary size of 10,000 individuals. The size of the bottleneck is varied in the table below. A slightly less severe bottleneck (1000 versus 550) was chosen for our analyses because it was more consistent with the observed genome-wide autosomal data.

| Diversity | Observed | Modeled diversity with varying bottlenecks | | |
| --- | --- | --- | --- | --- |
| | | 550 individuals | 1000 individuals | 2000 individuals |
| Autosome | 0.0563 | 0.04832 | 0.05633 | 0.06196 |
| chrX | 0.0365 | 0.03330 | 0.03992 | 0.04541 |
| chrY | 0.0024 | 0.00695 | 0.01008 | 0.01299 |
| mtDNA | 0.0147 | 0.00682 | 0.01005 | 0.01302 |

**Table S7. Comparing chromosome-wide SNPs.** In an effort to determine whether the Complete Genomics dataset is dramatically under-calling SNPs on the Y chromosome, we compared the total number of sites and SNPs in the set of sixteen male samples that overlap between the unrelated males in the Complete Genomics public dataset, and the 1000 genomes dataset: NA19700 (ASW), NA19703 (ASW), NA19834 (ASW), NA18501 (YRI), NA18504 (YRI), NA19020 (LWK), HG00731 (PUR), NA19735 (MXL), NA20509 (TSI), NA20510 (TSI), NA06994 (CEU), NA07357 (CEU), NA10851 (CEU), NA12889 (CEU), NA18558 (CHB), and NA18940 (JPT). We find that, whether we use no filtering (SNPs called on the Y in any 16 males, regardless of whether the sites were called in any other individual), or the same conservative filtering applied in the main manuscript (requiring that sites be called in all individuals analyzed), there are many more SNPs called in the Complete Genomics dataset, than in the 1000 genomes. This cannot be attributed to different amounts of sequences assayed, as both assay roughly 22Mb of sequence on the Y chromosome. Because the 1000 genomes does not report sites called for each individual, we report data from the one mask file they share, which is sites called across all individuals. For the Complete Genomics data we report both filters of the total number of sites assayed (either called in any individual, or requiring the site be called in all).

| Filtering | Complete Genomics | 1000 Genomes | Sites called in 1000 genomes, and not CG |
|---|---|---|---|
| SNPs in any 16 males | 13197 | 2136 | 213 (all singletons) |
| SNPs in all 16 males | 6236 | 252 | 12 (all singletons) |
| Total sites | 22848852*/20480009** | 22984529 | - |

*Sites in any individual
**Sites in all individuals

**Table S8. Parameters used in ms simulations.** For all simulations, African: size1 = 0.5, and European: size1 = 0.01, and size2 = 1. The models for Africans assume an expansion from 10,000 to 20,000 individuals, 4,000 generations ago, and for Europeans a bottleneck from 1,500 to 1,100 generations ago, from an ancestral size of 10,000 individuals, reduced to 1,000, then expanded back to 10,000 individuals. As the ratio of the effective number of males and females ($N_m/N_f$) varies, the effective population size of the autosomes (A) is held constant.

| Chr | $N_m/N_f$ | Africans $N_o$ | theta | rho | time1 | Europeans $N_o$ | theta | rho | time1 | time2 |
|---|---|---|---|---|---|---|---|---|---|---|
| A | 1 | 20000.00 | 1688.54 | 800.00 | 0.0500 | 10000.00 | 844.27 | 400.00 | 0.0275 | 0.0375 |
| X | 1 | 15000.00 | 991.19 | 400.00 | 0.0667 | 7500.00 | 495.60 | 200.00 | 0.0367 | 0.0500 |
| X | 0.9 | 15267.86 | 1008.89 | 407.14 | 0.0655 | 7633.93 | 504.45 | 203.57 | 0.0360 | 0.0491 |
| X | 0.75 | 15750.00 | 1040.75 | 420.00 | 0.0635 | 7875.00 | 520.38 | 210.00 | 0.0349 | 0.0476 |
| X | 0.5 | 16875.00 | 1115.09 | 450.00 | 0.0593 | 8437.50 | 557.55 | 225.00 | 0.0326 | 0.0444 |
| X | 0.38 | 17642.05 | 1165.78 | 470.45 | 0.0567 | 8821.02 | 582.89 | 235.23 | 0.0312 | 0.0425 |
| X | 0.25 | 18750.00 | 1238.99 | 500.00 | 0.0533 | 9375.00 | 619.50 | 250.00 | 0.0293 | 0.0400 |
| X | 0.1 | 20625.00 | 1362.89 | 550.00 | 0.0485 | 10312.50 | 681.45 | 275.00 | 0.0267 | 0.0364 |
| Y | 1 | 5000.00 | 5305.66 | - | 0.2000 | 2500.00 | 2652.83 | - | 0.1100 | 0.1500 |
| Y | 0.9 | 4750.00 | 5040.38 | - | 0.2105 | 2375.00 | 2520.19 | - | 0.1158 | 0.1579 |
| Y | 0.75 | 4375.00 | 4642.46 | - | 0.2286 | 2187.50 | 2321.23 | - | 0.1257 | 0.1714 |
| Y | 0.5 | 3750.00 | 3979.25 | - | 0.2667 | 1875.00 | 1989.62 | - | 0.1467 | 0.2000 |
| Y | 0.38 | 3450.00 | 3660.91 | - | 0.2899 | 1725.00 | 1830.45 | - | 0.1594 | 0.2174 |
| Y | 0.25 | 3125.00 | 3316.04 | - | 0.3200 | 1562.50 | 1658.02 | - | 0.1760 | 0.2400 |
| Y | 0.1 | 2750.00 | 2918.12 | - | 0.3636 | 1375.00 | 1459.06 | - | 0.2000 | 0.2727 |
| M | 1 | 5000.00 | 6.17 | - | 0.2000 | 2500.00 | 3.09 | - | 0.1100 | 0.1500 |
| M | 0.9 | 5277.78 | 6.51 | - | 0.1895 | 2638.89 | 3.26 | - | 0.1042 | 0.1421 |
| M | 0.75 | 5833.33 | 7.20 | - | 0.1714 | 2916.67 | 3.60 | - | 0.0943 | 0.1286 |
| M | 0.5 | 7500.00 | 9.26 | - | 0.1333 | 3750.00 | 4.63 | - | 0.0733 | 0.1000 |
| M | 0.38 | 9078.95 | 11.21 | - | 0.1101 | 4539.47 | 5.60 | - | 0.0606 | 0.0826 |
| M | 0.25 | 12500.00 | 15.43 | - | 0.0800 | 6250.00 | 7.71 | - | 0.0440 | 0.0600 |
| M | 0.1 | 27500.00 | 33.94 | - | 0.0364 | 13750.00 | 16.97 | - | 0.0200 | 0.0273 |

**Supplementary Figures**

**Figure S1. Divergence on the Y chromosome.** If we are overcorrecting for human-chimpanzee divergence on the Y chromosome, then we may under-estimate normalized Y diversity. To assess how different divergence corrections affect the normalized diversity estimate on the Y, we corrected raw Y diversity by divergence computed for the autosomes (Y_A), the X chromosome (Y_X), the Y chromosome (Y_Y) and the mtDNA (Y_mtDNA). In all cases, normalized diversity on the Y chromosome is still extremely low.

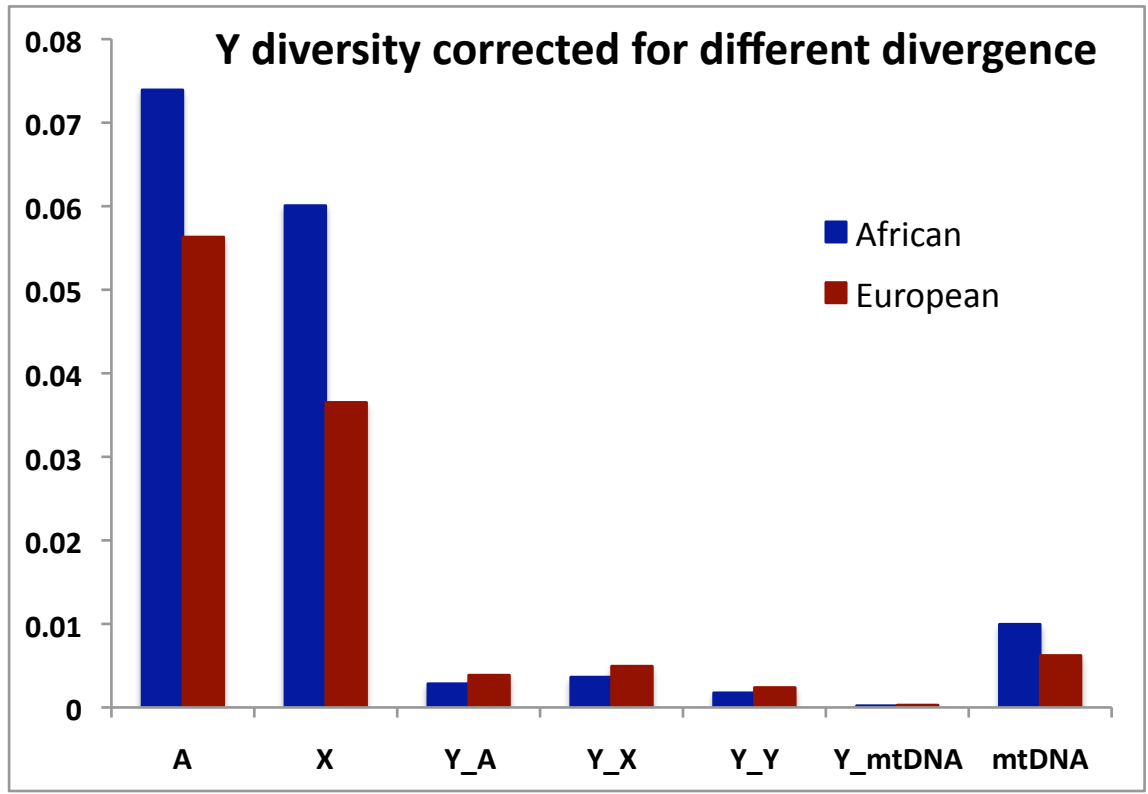

**Figure S2. Diploid versus haploid calling.** If Complete Genomics is under-calling variation on the Y chromosome due to its haploid nature, then we would expect it to also under-call variation on the X chromosome in males. So, we compared diversity on the X chromosome, when computed across the unrelated African and European males (as presented in the main text) with diversity on the X as computed in the unrelated African and European females, and find that there are no large reductions in diversity on the X, when using males, m, (where the X is haploid) versus females, f, (where the X is diploid).

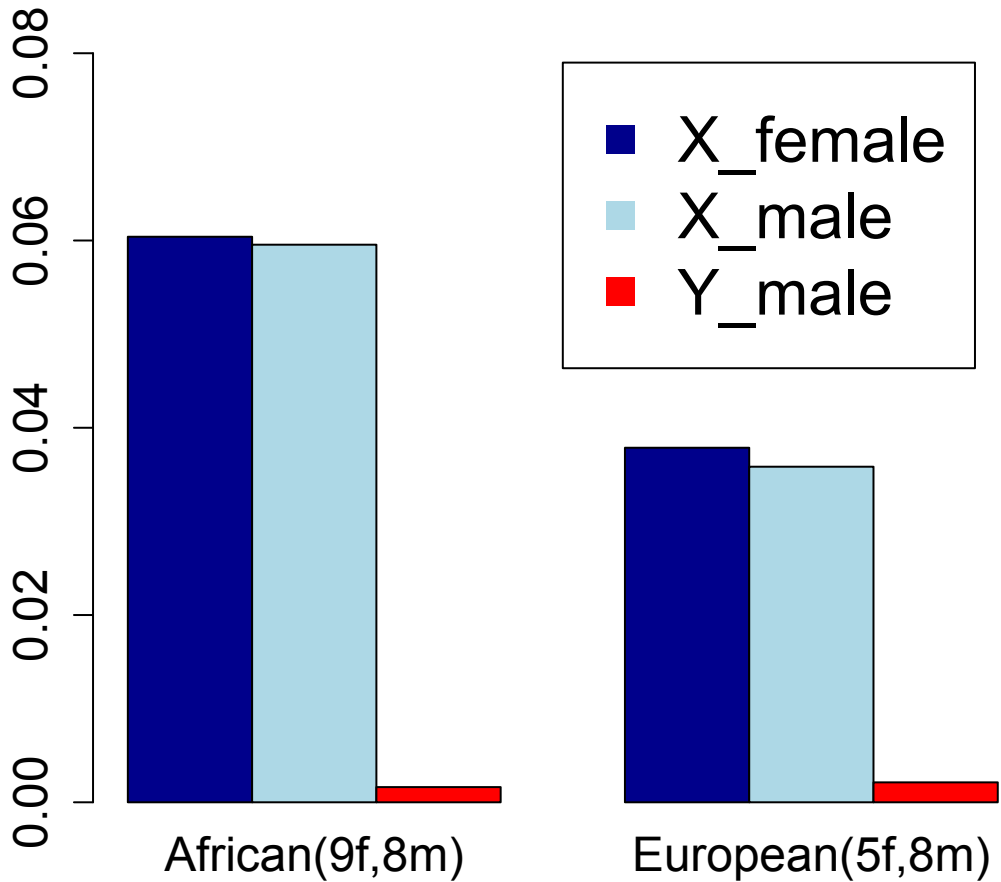

**Figure S3. Diversity across populations.** All major populations from the Complete Genomics data were initially analyzed (African, African without African Americans, Hispanic, European, East Asian and Indian), using both the male and the female samples for all chromosome regions (Autosomes, Chromosome X, Chromosome Y and mtDNA. Where possible we analyzed all individuals from the subpopulation regardless of gender (analyzing all autosomes from the populations available from the complete set of 54 individuals), or analyzed diversity on the X and mtDNA across all females from the subpopulation from the 26 unrelated females, or analyzed diversity on the X, mtDNA and Y across all males from the subpopulation from the 28 unrelated males. In every population, diversity on the Y was noticeably reduced.

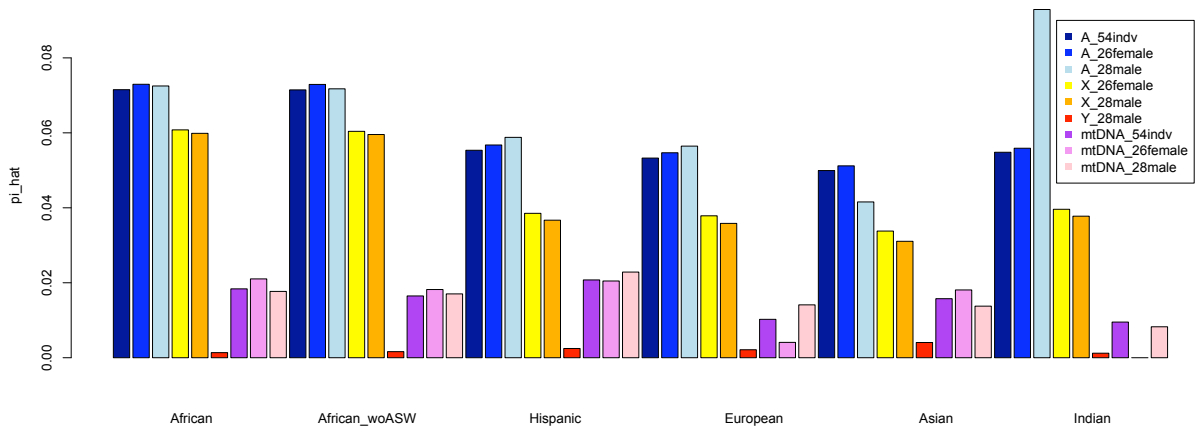

**Figure S4. Purifying selection on coding sites.** Models of selection acting only on coding sites cannot sufficiently reduce expected diversity on the Y chromosome. Under assumptions of equal sex ratios ($N_m/N_f = 1$), simulations using different mean selection coefficients (s), result in expectations of diversity that are significantly different from the observed values for both Africans and Europeans. Alternatively, under an assumption of an extremely reduced male effective population size, relative to females ($N_m/N_f = 0.38$), diversity is still significantly different from observations in Africans, but passes into a range of being consistent with observed diversity in Europeans. We fixed the shape parameter to 0.184 and performed simulations using mean selection coefficients ranging from 0.0001 to 0.09.

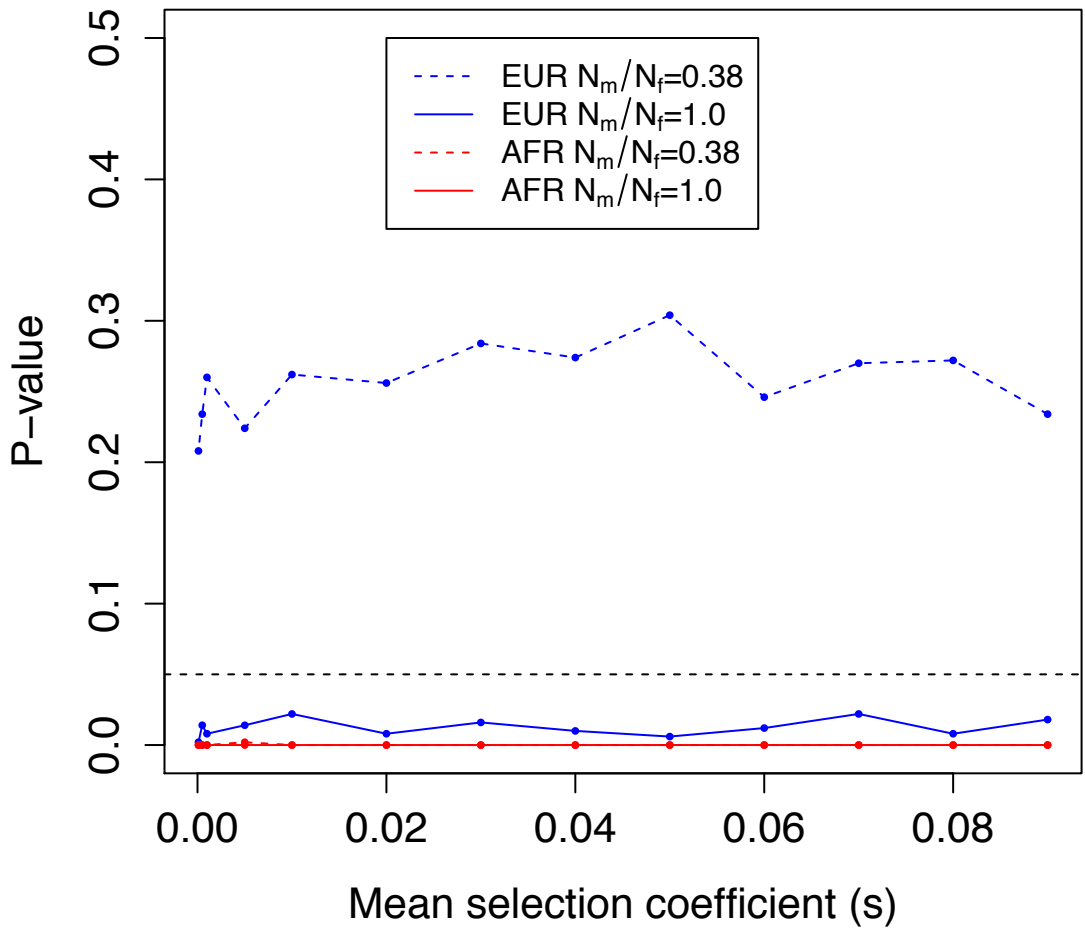

**Figure S5. Branch length comparisons.** Total tree length cannot easily be estimated from forward simulations with multiple sites under selection but can be approximated by the number of segregating sites. Comparisons are shown of the distribution of total branch lengths (in units of generations) for genealogies of 8 chromosomes from a population of size 10,000 diploids simulated under the standard neutral model without recombination using ms. The black curve denotes the actual distribution of total branch length for all of the 10,000 simulated genealogies. The remaining curves show the distribution of branch lengths estimated by dividing the number of segregating sites by $\mu_{sim}$. As $\mu_{sim}$ becomes large (>0.0075), the distribution of branch lengths inferred from the number of segregating sites approaches the true distribution. In other words, the majority of variance in the number of segregating sites is due to the variance of the total branch lengths, rather than the Poisson variance in the number of mutations conditional on the given genealogy. As such, the number of segregating sites in simulations can be taken as a reasonable proxy for the total branch lengths of the genealogy.

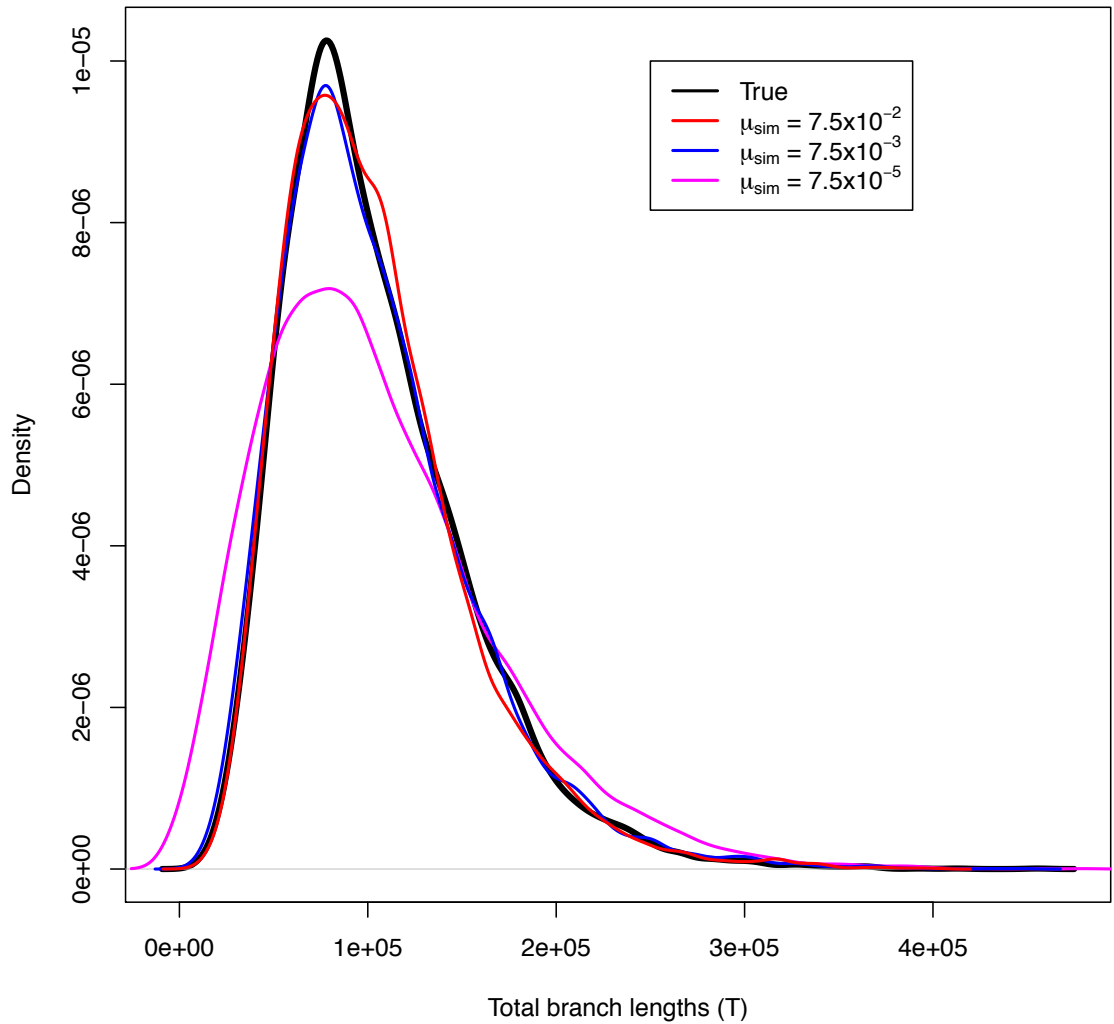

**Figure S6. Log-likelihood curves for the number of sites affected by purifying selection ($L$).** Blue curves are from the European population and red curves are from the Africans. Dotted lines denote sex-biased demography where $N_m/N_f = 0.1$, the dashed lines denote sex-biased demography where $N_m/N_f = 0.38$ while the solid lines represent equal sex ratios ($N_m/N_f = 1$). The horizontal dashed line denotes the asymptotic 95% confidence interval cutoff (1.92 log-likelihood units), such that values above this line result in estimates of diversity that are consistent with observations.

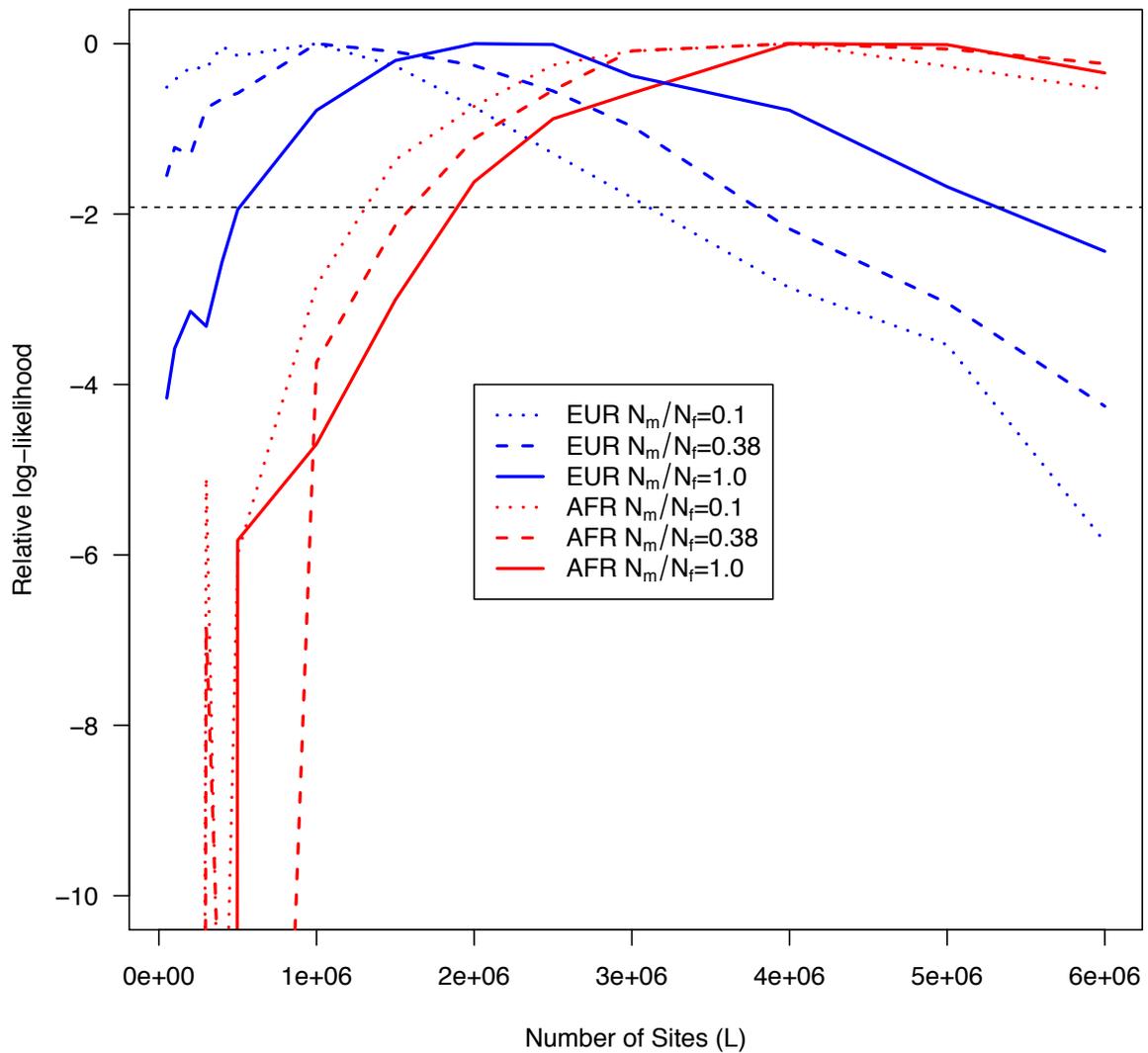

**Figure S7. Profile log-likelihood curves for the number of sites affected by purifying selection (*L*) when jointly estimating *L* and the mean strength of selection.** Blue curves are from the European population and red curves are from the Africans. The horizontal dashed line denotes the asymptotic 95% confidence interval cutoff (1.92 log-likelihood units), such that values above this line result in estimates of diversity that are consistent with observations.

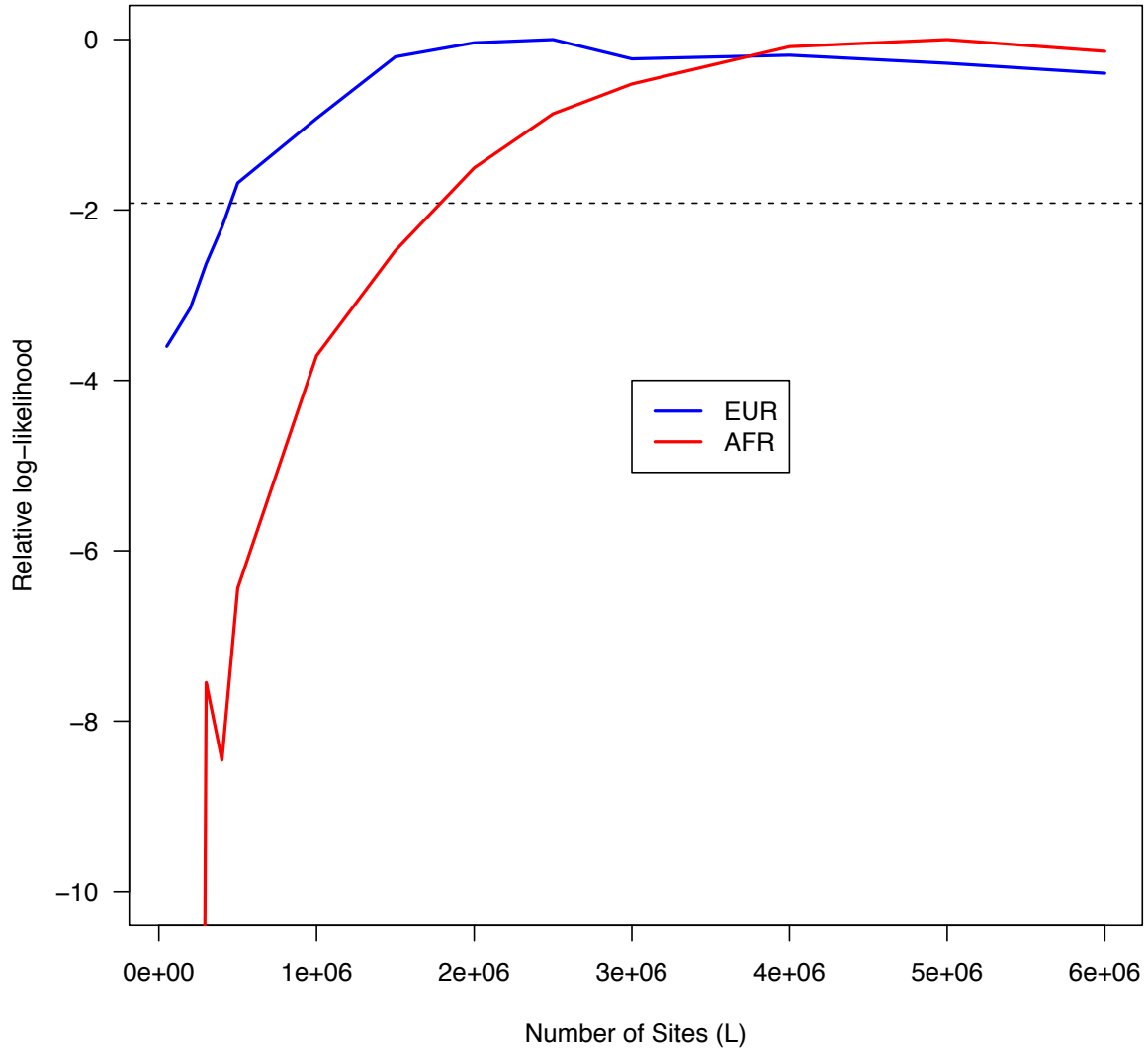

**Figure S8. Y chromosome site frequency spectrum.** The folded site frequency spectrum for Y chromosome SNPs across all 28 unrelated Y chromosomes in the Complete Genomics dataset is shown below. We include all unrelated males to limit reduce stochastic variation from analyzing only the 8 African and 8 Europeans of the main text. The abundance of low frequency SNPs is consistent with both positive selection and purifying selection. Given the very low total number of SNPs on the human Y chromosome, and the high divergence between human and chimpanzee, we only analyzed the folded site frequency spectrum.

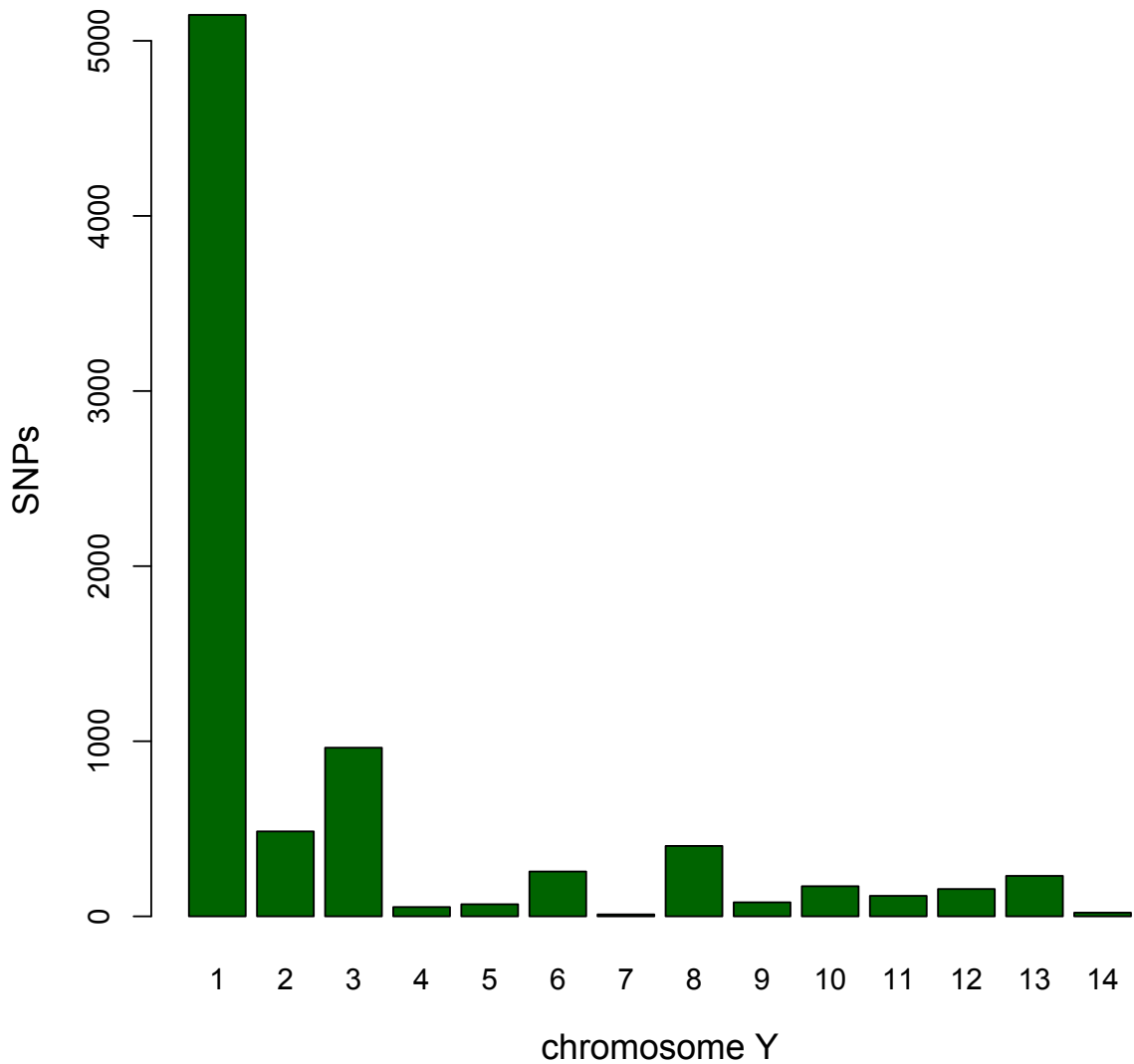

**Figure S9. Y chromosome haplotype tree.** A neighbor-joining tree was built for all unrelated Y haplotypes in the Complete Genomics dataset in phylip [50], then branch lengths were computed using a molecular clock in paml [51]. We include all 28 males to gain a higher resolution into the topology of the phylogenetic tree. There is not an overarching star phylogeny, which would be indicative of positive selection. However, one might conceive of a complex evolutionary history involving several instances of positive selection along different Y lineages that could result in the observed haplotype topology. The colors correspond to reported population ancestry: red names correspond to individuals reported to be of African or African American descent, yellow corresponds to individuals reported to be of Asian descent, blue corresponds to people reported to be of European descent, and green corresponds to people reported to be of Latino descent.

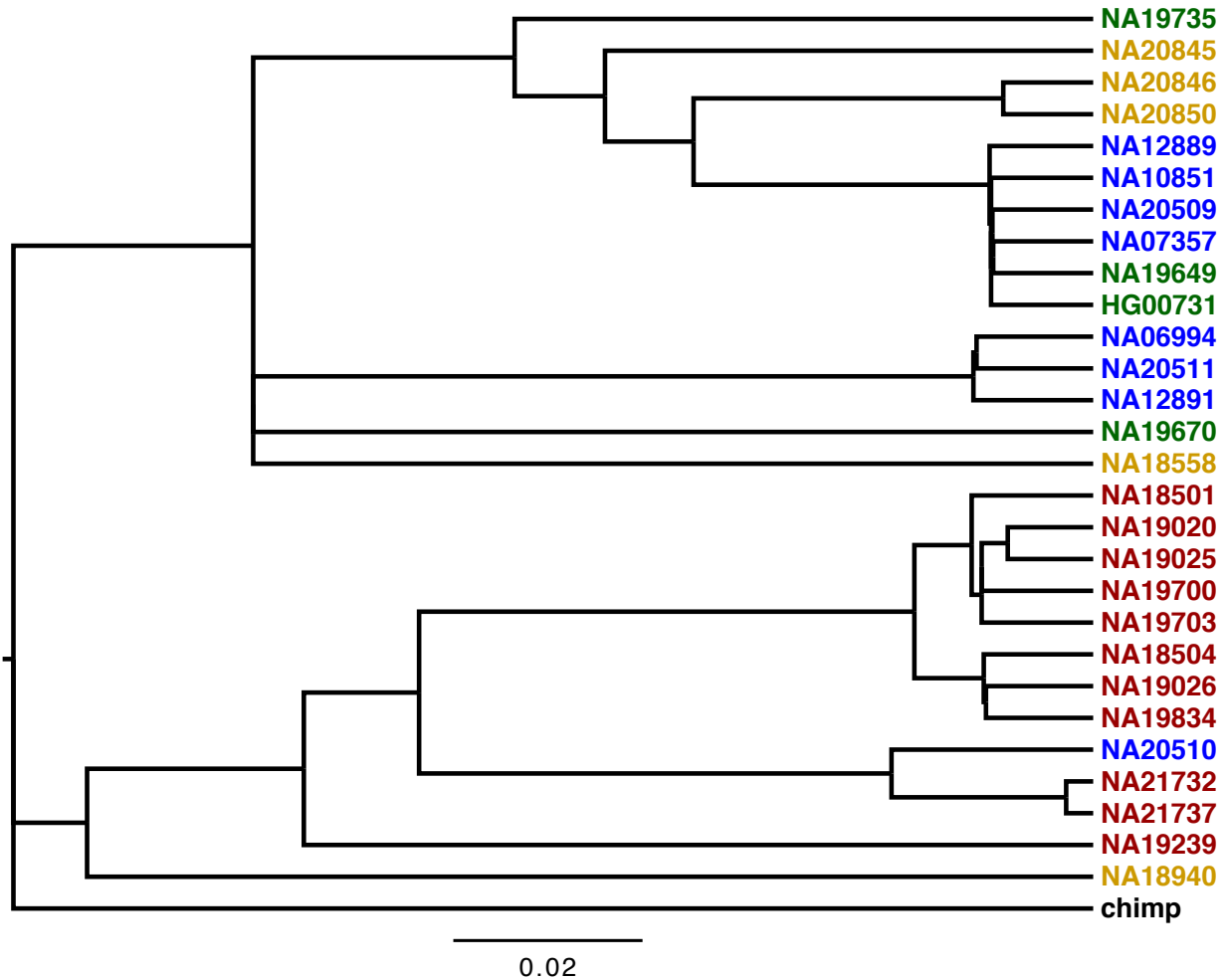